\documentstyle[12pt,epsf,citesort]{article}

\newcommand{\1}{{\sf 1 \!\! 1}}

\makeatletter
\@addtoreset{equation}{section}
\makeatother

\title{D-Theory: Field Quantization by Dimensional Reduction of Discrete
Variables}
\author{R.~Brower$^a$, S.~Chandrasekharan$^b$, S.~Riederer$^c$, and 
U.-J.~Wiese$^c$\footnote{on leave from MIT} \\ \\
$^a$ Department of Physics, Boston University \\ 
Boston, Massachusetts 02215, U.S.A. \\ \\
$^b$ Department of Physics, Duke University \\
Durham, North Carolina, Box 90305, U.S.A. \\ \\
$^c$ Institute for Theoretical Physics, Bern University \\
CH-3012 Bern, Switzerland}

\begin{document} 
\maketitle
\begin{abstract} \normalsize

D-theory is an alternative non-perturbative approach to quantum field theory
formulated in terms of discrete quantized variables instead of classical 
fields. Classical scalar fields are replaced by generalized quantum spins and 
classical gauge fields are replaced by quantum links. The classical fields of a
$d$-dimensional quantum field theory reappear as low-energy effective degrees 
of freedom of the discrete variables, provided the $(d+1)$-dimensional D-theory
is massless. When the extent of the extra Euclidean dimension becomes small 
in units of the correlation length, an ordinary $d$-dimensional quantum field 
theory emerges by dimensional reduction. The D-theory formulation of scalar
field theories with various global symmetries and of gauge theories with 
various gauge groups is constructed explicitly and the mechanism of dimensional
reduction is investigated.

\end{abstract}
 
\maketitle
 
\newpage

\section{Introduction}

Field theories are usually quantized by performing a path integral over
configurations of classical fields. This is the case both in perturbation 
theory and in Wilson's non-perturbative lattice formulation of quantum field
theory \cite{Wil74}. However, there is another form of quantization, which is 
well-known from quantum mechanics: a classical angular momentum vector can be 
replaced by a vector of Pauli matrices. The resulting quantum spin is described
by discrete variables $\pm 1/2$, while the original classical angular momentum 
vector represents a continuous degree of freedom. Here we apply this kind of 
quantization to field theory.

Of course, it is far from obvious that such a quantization procedure is 
equivalent to the usual one. For example, a single spin $1/2$ has the same
symmetry properties as a classical angular momentum vector, but it operates in 
a finite Hilbert space. How can the full Hilbert space of a quantum field 
theory be recovered when the classical fields are replaced by analogs of
quantum spins? Indeed, as we will see, this requires a specific dynamics, 
which, however, is generic in a wide variety of cases. This includes scalar
field theories as well as gauge theories. In these cases, a collective 
excitation of a large number of discrete variables acts as a classical field, 
in the same way as many spins $1/2$ can act as a classical angular momentum 
vector. In order to collect a large number of discrete variables, it turns out 
to be necessary to formulate the theory with an additional Euclidean dimension.
When the $(d+1)$-dimensional theory is massless, the classical fields emerge as
low-energy excitations of the discrete variables. Once the extent of the extra 
dimension becomes small in units of the correlation length, the desired 
$d$-dimensional quantum field theory emerges via dimensional reduction. 
{\em Dimensional} reduction of {\em discrete} variables is a generic phenomenon
that occurs in a variety of models, thus leading to an alternative 
non-perturbative formulation of quantum field theory which we call 
{\em D-theory} \cite{Cha97,Bro99,Wie99,Sch01}. Gauge theories of discrete 
quantum variables were first discussed by Horn \cite{Hor81} and later by Orland
and Rohrlich \cite{Orl90}. They were rediscovered and related to standard gauge
theories via dimensional reduction in \cite{Cha97}.

D-theory has several interesting features that go beyond Wilson's 
non-per\-tur\-ba\-tive lattice formulation of quantum field theory. For 
example, 
due to the use of discrete variables, the theory can be completely fermionized.
All bosonic fields can be written as pairs of fermionic constituents, which we 
call rishons. ``Rishon'' is Hebrew, means ``first'', and has been used as a
name for fermionic constituents of gauge bosons \cite{Har79}. In contrast to
traditional composite models, the rishons of D-theory propagate at the cut-off
scale and thus are not directly related to physical particles. The two indices 
of a bosonic matrix field --- for example, the two color indices of a gluon 
field matrix --- can be separated because they are carried by two different 
rishons. This may lead to new ways to attack the large $N$ limit of QCD and 
other interesting field theories \cite{Bae02}. D-theory is also attractive from
a computational point of view. Discrete variables are particularly well suited 
for numerical simulations using very powerful cluster algorithms. For example, 
in this way it has been possible to simulate the D-theory version of the 2-d 
$O(3)$ model at non-zero chemical potential \cite{Cha02}, which remains 
impossible using Wilson's approach.

In this paper, we concentrate on the algebraic structure of D-theory. It
is organized as follows. In section 2, D-theory is explained in the context of 
the $O(3)$ model. Section 3 contains the D-theory representation of real and 
complex vector and matrix fields. In particular, we construct basic building 
blocks that can be used in a variety of D-theory models. For example, the 
discrete quantum analog of an $O(N)$ symmetric real scalar field is a 
generalized quantum spin in the algebra of $SO(N+1)$, while a $U(N)$ symmetric 
complex scalar field is represented by an $SU(N+1)$ quantum spin. Similarly, 
the quantum link variables that arise in the D-theory formulation of gauge 
theories with $SO(N)$, $SU(N)$, and $Sp(N)$ gauge groups naturally live in the 
algebras $SO(2N)$, $SU(2N)$, and $Sp(2N)$. These algebraic structures provide 
the basis for the explicit construction of various models in section 4. This
includes $O(N)$, $U(N) \otimes U(N)$, and $CP(N)$ quantum spin models, as
well as $SO(N)$, $U(N)$, and $SU(N)$ quantum link models. The dimensional 
reduction to ordinary scalar field theories or gauge theories is discussed in 
section 5. In particular, formulas are derived for the finite correlation 
lengths that arise in the dimensionally reduced theory. Finally, section 6 
contains our conclusions.

\section{The $O(3)$ Model from D-Theory}
 
Let us illustrate the basic ideas behind D-theory in the simplest example
--- the 2-d $O(3)$ model, which we view as a Euclidean field theory in $1+1$
dimensions. Like QCD, this model is asymptotically free and has a 
non-perturbatively generated mass gap. In Wilson's formulation of lattice field
theory the model is formulated in terms of classical 3-component unit vectors 
$\vec s_x$ located on the sites $x$ of a quadratic lattice. The Euclidean 
action of the model is given by
\begin{equation}
S[\vec s] = - \sum_{x,\mu} \vec s_x \cdot \vec s_{x+\hat\mu},
\end{equation}
where $\hat\mu$ represents the unit vector in the $\mu$-direction. The theory
is quantized by considering the classical partition function
\begin{equation}
Z = \int {\cal D}\vec s \ \exp(- \frac{1}{g^2} S[\vec s]),
\end{equation}
which represents a path integral over all classical spin field configurations 
$[\vec s]$. Here $g$ is the coupling constant. Due to asymptotic freedom, the
continuum limit of the lattice model corresponds to $g \rightarrow 0$. In this
limit the correlation length $\xi \propto \exp(2 \pi/g^2)$ diverges 
exponentially. The strength of the exponential increase is given by the 1-loop 
$\beta$-function coefficient $1/2 \pi$ of the 2-d $O(3)$ model.

In contrast to the standard procedure, in D-theory one does not quantize by 
integrating over the classical field configurations $[\vec s]$. Instead, each 
classical vector $\vec s_x$ is replaced by a quantum spin operator $\vec S_x$ 
(a Pauli matrix for spin $1/2$) with the usual commutation relations
\begin{equation}
[S_x^i,S_y^j] = i \delta_{xy} \epsilon_{ijk} S_x^k.
\end{equation}
The classical action of the 2-d $O(3)$ model is replaced by the action operator
\begin{equation}
H = J \sum_{x,\mu} \vec S_x \cdot \vec S_{x+\hat\mu},
\end{equation}
which resembles the Hamilton operator of a quantum Heisenberg model. Here we 
restrict ourselves to antiferromagnets, i.e.\ to $J>0$. Ferromagnets have a 
conserved order parameter, and therefore require a special treatment. Like the 
classical action $S[\vec s]$, the action operator $H$ is invariant under global
$SO(3)$ transformations. In quantum mechanics this follows from 
$[H,\vec S] = 0$, where
\begin{equation}
\vec S = \sum_x \vec S_x
\end{equation}
is the total spin. D-theory is defined by the quantum partition function
\begin{equation}
Z = \mbox{Tr} \exp(- \beta H).
\end{equation}
The trace is taken in the Hilbert space, which is a direct product of the
Hilbert spaces of individual spins. It should be noted that D-theory can be
formulated with any value of the spin, not only with spin $1/2$.

At this point, we have replaced the 2-d $O(3)$ model, formulated in terms of 
classical fields $\vec s_x$, by a 2-d system of quantum spins $\vec S_x$ with 
the same symmetries. The inverse ``temperature'' $\beta$ of the quantum system 
can be viewed as the extent of an additional third dimension. In the condensed
matter interpretation of the 2-d quantum spin system this dimension would be 
Euclidean time. In D-theory, however, Euclidean time is already part of the 2-d
lattice. Indeed, as we will see, the additional Euclidean dimension ultimately 
disappears via dimensional reduction. The 2-d antiferromagnetic spin $1/2$ 
quantum Heisenberg model has very interesting properties. It describes the 
undoped precursor insulators of high-temperature superconductors --- materials 
like $\mbox{La}_2\mbox{CuO}_4$ --- whose ground states are N\'eel ordered with 
a spontaneously generated staggered magnetization. Indeed, there is 
overwhelming numerical evidence that the ground state of the 2-d 
antiferromagnetic spin $1/2$ quantum Heisenberg model exhibits long-range order
\cite{Bar91,Wie94,Bea96}. The same is true for higher spins, and thus the 
following discussion applies equally well to all spin values. In practice, 
however, the smallest spin $1/2$ is most interesting, because it allows us to 
represent the physics of the 2-d $O(3)$ model in the smallest possible Hilbert 
space.

Formulating the 2-d quantum model as a path integral in the extra dimension 
results in a 3-d $SO(3)$-symmetric classical model. At zero temperature of the 
quantum system we are in the infinite volume limit of the corresponding 3-d 
model. The N\'eel order of the ground state of the 2-d quantum system implies 
that the corresponding 3-d classical system is in a broken phase, in which only
an $SO(2)$ symmetry remains intact. As a consequence of Goldstone's theorem, 
two massless bosons arise --- in this case two antiferromagnetic magnons (or 
spin-waves). Using chiral perturbation theory one can describe the magnon 
dynamics at low energies \cite{Leu90}. The Goldstone bosons are represented by 
fields in the coset $SO(3)/SO(2) = S^2$, which resembles a 2-dimensional 
sphere. Consequently, the magnons are described by 3-component unit vectors 
$\vec s$ --- the same fields that appear in the original 2-d $O(3)$ model. 
Indeed, due to spontaneous symmetry breaking, the collective excitations of 
many discrete quantum spin variables form an effective continuous classical 
field $\vec s$. This is one of the main dynamical ingredients of D-theory. 

Another ingredient is dimensional reduction, to which we now turn. To lowest 
order in chiral perturbation theory, the effective action of the Goldstone 
bosons takes the form
\begin{equation}
\label{spinaction}
S[\vec s] = \int_0^\beta dx_3 \int d^2x \ \frac{\rho_s}{2}
\left(\partial_\mu \vec s \cdot \partial_\mu \vec s + 
\frac{1}{c^2} \partial_3 \vec s \cdot \partial_3 \vec s\right).
\end{equation}
Here $c$ and $\rho_s$ are the spin-wave velocity and the spin stiffness. Note
that $\mu$ extends over the physical space-time indices 1 and 2 only. The 2-d 
quantum system at finite temperature corresponds to a 3-d classical model with 
finite extent $\beta$ in the extra dimension. For massless particles --- 
i.e.\ in the presence of an infinite correlation length $\xi$ --- the finite
temperature system appears dimensionally reduced to two dimensions, because 
$\beta \ll \xi$. In two dimensions, however, the Mermin-Wagner-Coleman theorem 
prevents the existence of interacting massless Goldstone bosons, and, indeed, 
the 2-d $O(3)$ model has a non-perturbatively generated mass gap. Hasenfratz 
and Niedermayer used a block spin renormalization group transformation to map 
the 3-d $O(3)$ model with finite extent $\beta$ to a 2-d lattice $O(3)$ model 
\cite{Has91}. One averages the 3-d field over volumes of size $\beta$ in the 
third direction and $\beta c$ in the two space-time directions. Due to the 
large correlation length, the field is essentially constant over these blocks. 
The averaged field is defined at the block centers, which form a 2-d lattice of
spacing $\beta c$. Note that this lattice spacing is different from the lattice
spacing of the original quantum antiferromagnet. The effective action of the 
averaged field defines a 2-d lattice $O(3)$ model, formulated in Wilson's 
framework. Using chiral perturbation theory, Hasenfratz and Niedermayer 
expressed its coupling constant as
\begin{equation}
1/g^2 = \beta \rho_s - \frac{3}{16 \pi^2 \beta \rho_s} + 
{\cal O}(1/\beta^2 \rho_s^2).
\end{equation}
Using the 3-loop $\beta$-function of the 2-d $O(3)$ model together with its
exact mass gap \cite{Has90}, they also extended an earlier result of 
Chakravarty, Halperin, and Nelson \cite{Cha89} for the correlation 
length of the quantum antiferromagnet to
\begin{equation}
\xi = \frac{ec}{16 \pi \rho_s} \exp(2 \pi \beta \rho_s)
\left[1 - \frac{1}{4 \pi \beta \rho_s} + {\cal O}(1/\beta^2 \rho_s^2)\right].
\end{equation}
Here $e$ is the base of the natural logarithm. The above equation resembles
the asymptotic scaling behavior of the 2-d classical $O(3)$ model. In fact, one
can view the 2-d antiferromagnetic quantum Heisenberg model in the zero 
temperature limit as a regularization of the 2-d $O(3)$ model. It is remarkable
that this D-theory formulation is entirely discrete, even though the model is 
usually formulated with a continuous classical configuration space.

The dimensionally reduced effective 2-d theory is a Wilsonian lattice theory 
with lattice spacing $\beta c$. The continuum limit of that theory is reached 
as $g^2 = 1/\beta \rho_s \rightarrow 0$, and hence as the extent $\beta$ of the
extra dimension becomes large. Still, in physical units of the correlation 
length, the extent $\beta \ll \xi$ becomes negligible in this limit, and hence 
the theory undergoes dimensional reduction to two dimensions. In the continuum 
limit, the lattice spacing $\beta c$ of the effective 2-d Wilsonian lattice 
$O(3)$ model becomes large in units of the microscopic lattice spacing of the 
quantum spin system. Therefore, D-theory introduces a discrete substructure 
underlying Wilson's lattice theory. This substructure is defined on a very fine
microscopic lattice. In other words, D-theory regularizes quantum fields at 
much shorter distance scales than the ones considered in Wilson's formulation. 

The additional microscopic structure may provide new insight into the 
long-distance continuum physics. In the context of the quantum Heisenberg 
model, the microscopic substructure is due to the presence of electrons hopping
on a crystal lattice. After all, the spin waves of a quantum antiferromagnet 
are just collective excitations of the spins of many electrons. In the same 
way, gluons appear as collective excitations of rishons hopping on the 
microscopic lattice of the corresponding quantum link model for QCD. In that 
case, the lattice is unphysical and just serves as a regulator. However, even 
if the rishons propagate only at the cut-off scale, they may still be useful 
for understanding the physics in the continuum limit. Let us illustrate the 
rishon idea in the context of the quantum Heisenberg model. Then the rishons 
can be identified with physical electrons. In fact, the quantum spin operator 
at a lattice site $x$,
\begin{equation}
\vec S_x = \frac{1}{2} \sum_{i,j} c_x^{i\dagger} \vec \sigma_{ij} c_x^j,
\end{equation}
can be expressed in terms of Pauli matrices $\vec \sigma$ and electron creation
and annihilation operators $c_x^{i\dagger}$ and $c_x^i$ ($i,j \in \{1,2\}$)
with the usual anti-commutation relations
\begin{equation}
\{c_x^{i\dagger},c_y^{j\dagger}\} = \{c_x^i,c_y^j\} = 0, \ 
\{c_x^i,c_y^{j\dagger}\} = \delta_{x,y} \delta_{ij}.
\end{equation}
It is straightforward to show that $\vec S_x$, constructed in this way, has the
correct commutation relations. In fact, the commutation relations are also
satisfied when the rishons are quantized as bosons. It should be noted that the
total number of rishons at each site $x$ is a conserved quantity, because the 
local rishon number operator ${\cal N}_x = \sum_i c_x^{i\dagger} c_x^i$ 
commutes with the Hamiltonian. In fact, fixing the number of rishons is 
equivalent to selecting a value for the spin, i.e.\ to choosing an irreducible 
representation.

The discrete nature of the D-theory degrees of freedom allows the application 
of very efficient cluster algorithms. The quantum Heisenberg model, for 
example, can be treated with a loop cluster algorithm \cite{Eve93,Wie94}. 
Defining the path integral for discrete quantum systems does not even require 
discretization of the additional Euclidean dimension. This observation has led 
to a very efficient loop cluster algorithm operating directly in the continuum 
of the extra dimension \cite{Bea96}. This algorithm, combined with a 
finite-size scaling technique, has been used to study the correlation length
of the Heisenberg model up to $\xi \approx 350000$ lattice spacings 
\cite{Bea98}. In this
way the analytic prediction of Hasenfratz and Niedermayer --- and hence the 
scenario of dimensional reduction --- has been verified. This shows explicitly 
that the 2-d O(3) model can be investigated very efficiently using D-theory, 
i.e.\ by simulating the $(2+1)$-d path integral for the 2-d quantum Heisenberg 
model. In this case, the numerical effort is compatible to simulating the 2-d 
$O(3)$ model directly with the Wolff cluster algorithm \cite{Wol89}. However,
D-theory allows us to simulate the 2-d $O(3)$ model even at non-zero chemical
potential \cite{Cha02}, which has not been possible with traditional methods.
For most other lattice models --- for example, for gauge theories --- despite 
numerous attempts, no efficient cluster algorithm has been found in Wilson's 
formulation. If an efficient cluster algorithm can be constructed for the 
D-theory formulation, it would allow simulations much more accurate than the 
ones presently possible.

The exponential divergence of the correlation length is due to the asymptotic
freedom of the 2-d $O(3)$ model. Hence, one might expect that the above 
scenario of dimensional reduction is specific to $d = 2$. As we will see now,
dimensional reduction also occurs in higher dimensions, but in a slightly
different way. Let us consider the antiferromagnetic quantum Heisenberg model 
on a $d$-dimensional lattice with $d > 2$. Then, again, the ground state has a
broken symmetry, and the low energy excitations of the system are two 
massless magnons. The effective action of chiral perturbation theory is the 
same as before, except that the integration now extends over a 
higher-dimensional space-time. Again, at an infinite extent $\beta$ of the 
extra dimension, one has an infinite correlation length. Thus, once $\beta$ 
becomes finite, the extent of the extra dimension is negligible compared to the
correlation length, and the theory undergoes dimensional reduction to $d$ 
dimensions. However, in contrast to the $d = 2$ case, now there is no reason 
why the Goldstone bosons should pick up a mass after dimensional reduction. 
Consequently, the correlation length remains infinite and we end up in a 
$d$-dimensional phase with broken symmetry. When the extent $\beta$ is reduced 
further, we eventually reach the symmetric phase, in which the correlation 
length is finite. The transition between the broken and symmetric phase is 
known to be of second order. Thus, approaching the phase transition from the 
symmetric phase at low $\beta$ also leads to a divergent correlation length, 
and hence, again, to dimensional reduction. Thus, the universal continuum 
physics of $O(3)$ models in any dimension $d \geq 2$ is naturally contained in 
the framework of D-theory.

Still, the $d = 1$ case requires a separate discussion. The behavior of quantum
spin chains depends crucially on the value of the spin. Haldane has conjectured
that 1-d antiferromagnetic $O(3)$ quantum spin chains with integer spins have a
mass gap, while those with half-integer spins are gapless \cite{Hal83}. This
conjecture is by now verified in great detail. For example, the spin $1/2$ 
antiferromagnetic Heisenberg chain has been solved by the Bethe ansatz, and 
indeed turns out to have no mass gap \cite{Bet31}. The same has been shown 
analytically for all half-integer spins \cite{Lie61}. On the other hand, there 
is strong numerical evidence for a mass gap in spin 1 and spin 2 systems 
\cite{Bot84}. Hence, only for half-integer spins the $(1+1)$-dimensional 
D-theory with an infinite extent $\beta$ in the second direction has an 
infinite correlation length. The low-energy effective theory for this system is
the 2-d $O(3)$ model at vacuum angle $\theta = \pi$, i.e.
\begin{equation}
S[\vec s] = \int_0^\beta dx_1 \int dx_2 \ \left[\frac{1}{2 g^2}
\left(\partial_1 \vec s \cdot \partial_1 \vec s + 
\frac{1}{c^2} \partial_2 \vec s \cdot \partial_2 \vec s\right) +
\frac{i \theta}{4 \pi} \vec s \cdot 
\left(\partial_1 \vec s \times \partial_2 \vec s\right)\right],
\end{equation}
as conjectured by Haldane \cite{Hal83}. This model is in the universality class
of a 2-d conformal field theory --- the $k=1$ Wess-Zumino-Novikov-Witten model 
\cite{Nov81} --- as was first argued by Affleck \cite{Aff88}.  Indeed, it has 
been shown numerically that the mass gap of the 2-d $O(3)$ model (which is 
present at $\theta \neq \pi$) disappears at $\theta = \pi$ \cite{Bie95}. The 
simulation of the 2-d $O(3)$ model at $\theta = \pi$ is extremely difficult due
to the complex action. Still, it is possible using the Wolff cluster algorithm 
combined with an appropriate improved estimator for the topological charge 
distribution. It is remarkable that no complex action arises in the D-theory 
formulation of this problem, and the simulation is hence much simpler. When the
above model is dimensionally reduced by making the extent $\beta$ of the extra
dimension finite, the topological term disappears, because $\partial_t \vec s$ 
is then negligible due to the large correlation length. Using the same 
renormalization group argument as before, one obtains a 1-d $O(3)$ Wilsonian 
lattice model with effective coupling constant $\beta/g^2$. In the continuum 
limit $\beta \rightarrow \infty$ this model describes the quantum mechanics of 
a particle moving on the sphere $S^2$. Hence, D-theory even works for $d = 1$, 
however, only when formulated with half-integer spins. For integer spins there 
is no infinite correlation length, and hence one basic dynamical ingredient,
necessary for the success of D-theory, is missing. Still, in the classical 
limit of large integer spin $S$ the correlation length increases as 
$\xi \propto \exp(\pi S)$. Indeed, one could reach the continuum limit of the 
2-d $O(3)$ model with $1/g^2 = S/2$ in this way. However, this is not in the 
spirit of D-theory,  because one then effectively works with classical fields 
again.

So far, we have seen that D-theory naturally contains the continuum physics
of $O(3)$ models in any dimension. This alone would be interesting. However,
as we will see, D-theory is much more general. It can be extended to other
scalar field theories and to gauge theories with various symmetries and in
various space-time dimensions.

\section{D-Theory Representation of Basic Field Variables}

As we have seen, in the low-temperature limit the 2-d spin 1/2 quantum 
Heisenberg model provides a D-theory regularization for the 2-d 
$O(3)$-symmetric continuum field theory. In that case, a vector of Pauli 
matrices replaces the 3-component unit-vector of a classical field 
configuration. Here, this structure is generalized to other fields. 

\subsection{Real Vectors}

It is not obvious how the $N$-component unit-vectors of an $O(N)$ model should 
be represented in D-theory. An important hint comes from the quantum XY model
which has an $SO(2)$ symmetry. In that case, the Hamilton operator takes the 
form $H = J \sum_{x,\mu} (S^1_x S^1_{x+\hat\mu} + S^2_x S^2_{x+\hat\mu})$. The 
2-component unit vector $(s^1,s^2)$ of the classical XY model has been replaced
by the first two components of a quantum spin $(S^1,S^2)$ which indeed form a 
vector under $SO(2)$. This has a natural generalization to higher $N$. Let us 
consider the $(N+1)N/2$ generators of $SO(N+1)$. Among them, $N$ generators 
$S^i = S^{0i}$ ($i \in \{1,2,...,N\}$) transform as a vector under $SO(N)$ and
the remaining $N(N-1)/2$ generators $S^{ij}$ generate the subgroup $SO(N)$. In
other words, in the subgroup decomposition $SO(N+1) \supset SO(N)$ the adjoint
representation of $SO(N+1)$ decomposes as
\begin{equation}
\{\frac{(N+1)N}{2}\} = \{\frac{N(N-1)}{2}\} \oplus \{N\}.
\end{equation}
The commutation relations of the group $SO(N+1)$ take the form
\begin{eqnarray}
&&[S^i,S^j] = i S^{ij}, \
[S^i,S^{jk}] = i (\delta_{ik} S^j - \delta_{ij} S^k), \nonumber \\
&&[S^{ij},S^{kl}] = i (\delta_{il} S^{kj} + \delta_{ik} S^{jl} +
\delta_{jk} S^{li} + \delta_{jl} S^{ik}).
\end{eqnarray}
Just as in the quantum XY model, one works with an $SO(N+1)$ algebra, although 
the symmetry of the model is only $SO(N)$. It should be noted that for $N = 3$ 
the above construction does not reduce to the quantum Heisenberg model which 
is formulated in terms of the three generators of $SO(3)$. Instead, it yields 
another $SO(3)$-invariant quantum spin model expressed in terms of the six 
generators of $SO(4)$. In fact, the Heisenberg model construction is special 
and has no natural generalization to $O(N)$ models. The above commutation 
relations of $SO(N+1)$ can be represented by
\begin{equation}
S^i = S^{0i} = -i (c^0 c^i - c^i c^0), \ S^{ij} = -i (c^i c^j - c^j c^i).
\end{equation}
In this case, the rishon operators $c^0 = c^{0 \dagger}$ and 
$c^i = c^{i \dagger}$ are Hermitean and obey the anticommutation relations
\begin{equation}
\{c^0,c^i\} = 0, \ \{c^i,c^j\} = \delta_{ij}.
\end{equation}
Note that the Clifford algebra of these ``Majorana'' rishons can be represented
by ordinary $\gamma$-matrices.

\subsection{Real Matrices}

Spin models with an $SO(N)_L \otimes SO(N)_R$ symmetry are formulated in terms 
of classical real $O(N)$ matrix fields. Similarly, in $SO(N)$ lattice gauge 
theory one deals with real valued classical parallel transporter matrices $o$ 
which transform appropriately under $SO(N)_L \otimes SO(N)_R$ gauge 
transformations on the left and on the right. This symmetry is
generated by $N(N-1)$ Hermitean operators. In D-theory, the real valued 
classical matrix $o$ is replaced by an $N \times N$ matrix $O$ whose elements 
are Hermitean operators. Altogether, this gives $N(N-1) + N^2 = N(2N-1)$
generators --- the total number of generators of $SO(2N)$. The corresponding
subgroup decomposition $SO(2N) \supset SO(N)_L \otimes SO(N)_R$ yields 
\begin{equation}
\{N(2N-1)\} = \{\frac{N(N-1)}{2},1\} \oplus \{1,\frac{N(N-1)}{2}\} \oplus 
\{N,N\}. 
\end{equation}
Again, it is straightforward to show that the following rishon representation 
generates the algebra of $SO(2N)$
\begin{equation}
O^{ij} = -i (c_+^i c_-^j - c_-^j c_+^i), \ 
L^{ij} = -i (c_+^i c_+^j - c_+^i c_+^j), \
R^{ij} = -i (c_-^i c_-^j - c_-^i c_-^j).
\end{equation}
There are two sets of Hermitean ``Majorana'' rishons, $c_+^i = c_+^{i \dagger}$
and $c_-^i = c_-^{i \dagger}$, associated with the left and right $SO(N)$ 
symmetries generated by $\vec L$ and $\vec R$. They obey the anticommutation 
relations
\begin{equation}
\{c_+^i,c_+^j\} = \delta_{ij}, \ \{c_-^i,c_-^j\} = \delta_{ij}, \
\{c_+^i,c_-^j\} = 0.
\end{equation}

\subsection{Complex Vectors}

We have seen how to represent a real $N$-component vector $s$ in D-theory.
It is simply replaced by an $N$-component vector of Hermitean generators $S^i$ 
of $SO(N+1)$. In $CP(N-1)$ models, classical $N$-component complex vectors $z$ 
arise. We will now discuss their representation in D-theory. The symmetry group
of a $CP(N-1)$ model is $U(N)$ which has $N^2$ generators. In D-theory the 
complex components $z^i$ are represented by $2N$ Hermitean operators --- $N$ 
for the real and $N$ for the imaginary parts. Hence, the total number of 
generators is $N^2 + 2 N = (N+1)^2 - 1$ --- the number of generators of 
$SU(N+1)$. In this case, the subgroup decomposition $SU(N+1) \supset SU(N) 
\otimes U(1)$ takes the form
\begin{equation}
\{(N+1)^2 - 1\} = \{N^2 - 1\} \oplus \{1\} \oplus \{N\} \oplus
\{\overline{N}\}.
\end{equation}
A rishon representation of the $SU(N+1)$ algebra is given by
\begin{equation}
Z^i = c^{0 \dagger} c^i, \ 
\vec G = \sum_{ij} c^{i \dagger} \vec \lambda_{ij} c^j, \ 
G = \sum_i c^{i \dagger} c^i.
\end{equation}
In this case, we use ``Dirac'' rishons $c^0, c^{0 \dagger}, c^i, c^{i \dagger}$
with the usual anti-commutation relations. Here $\vec \lambda$ is the vector of
Gell-Mann matrices for $SU(N)$ which obeys 
$[\lambda^a,\lambda^b] = 2 i f_{abc} \lambda^c$ as well as 
$\mbox{Tr}(\lambda^a \lambda^b) = 2 \delta_{ab}$. The quantum operator $Z^i$ 
replaces the classical variable $z^i$, $\vec G$ is a vector of $SU(N)$ 
generators obeying $[G^a,G^b] = 2 i f_{abc} G^c$, and $G$ is a $U(1)$ 
generator.

\subsection{Complex Matrices}

Chiral spin models with a global $SU(N)_L \otimes SU(N)_R \otimes U(1)$ 
symmetry as well as $U(N)$ and $SU(N)$ lattice gauge theories are formulated in
terms of classical complex $U(N)$ matrix fields. The corresponding symmetry 
transformations are generated by $2(N^2-1)+1$ Hermitean operators. In D-theory 
a classical complex valued matrix $u$ is replaced by a matrix $U$ whose 
elements are non-commuting operators. The elements of the matrix $U$ are 
described by $2N^2$ Hermitean generators --- $N^2$ representing the real part 
and $N^2$ representing the imaginary part of the classical matrix $u$. 
Altogether, we thus have $2(N^2-1)+1+2N^2 = 4N^2-1$ generators --- the number 
of generators of $SU(2N)$. The corresponding subgroup decomposition $SU(2N) 
\supset SU(N)_L \otimes SU(N)_R \otimes U(1)$ takes the form 
\begin{equation}
\{4N^2 - 1\} = \{N^2 - 1,1\} \oplus \{1,N^2 - 1\} \oplus \{1,1\} \oplus 
\{N,\overline{N}\} \oplus \{\overline{N},N\}.
\end{equation}
A rishon representation of the $SU(2N)$ algebra is given by
\begin{equation}
U^{ij} = c_+^{i \dagger} c_-^j, \ 
\vec L = \sum_{ij} c_+^{i \dagger} \vec \lambda_{ij} c_+^j, \
\vec R = \sum_{ij} c_-^{i \dagger} \vec \lambda_{ij} c_-^j, \ 
T = \sum_i (c_+^{i \dagger} c_+^i - c_-^{i \dagger} c_-^i).
\end{equation}
There are two sets of ``Dirac'' rishons, $c_+^i$ and $c_-^i$, again associated 
with the left and right $SU(N)$ symmetries generated by $\vec L$ and $\vec R$. 
As before, $T$ is a $U(1)$ generator. For example, the above structure is used 
in $U(N)$ and $SU(N)$ quantum link models in which the elements of Wilson's 
classical parallel transporter matrices are replaced by non-commuting 
operators.

\subsection{Symplectic, Symmetric, and Anti-Symmetric Complex Tensors}

The third main sequence of Lie groups (besides $SO(N)$ and $SU(N)$) are the
symplectic groups $Sp(N)$. The group $Sp(N)$ is a subgroup of $SU(2N)$ whose
elements $g$ obey the additional constraint $g^* = J g J^\dagger$. The real 
skew-symmetric matrix $J$ obeys $J^2 = - \1$. It is interesting to ask how 
$Sp(N)$ gauge theories can be formulated in D-theory. In fact, this is 
completely analogous to the $SO(N)$ and $SU(N)$ cases. The 
$Sp(N)_L \otimes Sp(N)_R$ symmetry transformations are generated by $N(2N+1)$ 
Hermitean operators on the left and $N(2N+1)$ operators on the right. In 
D-theory, the $(2N)^2$ elements of an $Sp(N)$ symplectic matrix are described 
by $4N^2$ Hermitean operators. Altogether, we thus have $2N(2N+1) + 4N^2 = 
2N(4N+1)$ generators --- the number of generators of $Sp(2N)$. The 
corresponding subgroup decomposition $Sp(2N) \supset Sp(N)_L \otimes Sp(N)_R$ 
takes the form 
\begin{equation}
\{2N(4N+1)\} = \{N(2N+1),1\} \oplus \{1,N(2N+1)\} \oplus \{2N,2N\}.
\end{equation}

Other useful building blocks for D-theory models are symmetric ($S^T = S$) and 
anti-symmetric ($A^T = - A$) complex tensors which transform as
\begin{equation}
S' = g S g^T, \ A' = g A g^T,
\end{equation}
under $SU(N)$ transformations. In D-theory, the $N(N+1)/2$ elements of a
complex symmetric tensor are represented by $N(N+1)$ Hermitean operators. In
addition, there are again $N^2 - 1$ $SU(N)$ and one $U(1)$ generators. Hence,
altogether, there are $N(N+1) + N^2 = N(2N+1)$ generators --- namely those of
$Sp(N)$. In this case, the subgroup decomposition is $Sp(N) \supset SU(N) 
\otimes U(1)$ and it takes the form
\begin{equation}
\{N(2N+1)\} = \{N^2 - 1\} \oplus \{1\} \oplus \{\frac{N(N+1)}{2}\} \oplus
\{\overline{\frac{N(N+1)}{2}}\}.
\end{equation}
Similarly, the $N(N-1)/2$ elements of a complex anti-symmetric tensor are 
represented by $N(N-1)$ Hermitean operators. Again, there are also $N^2 - 1$ 
generators of $SU(N)$ and one $U(1)$ generator. In total, there are 
$N(N-1) + N^2 = N(2N-1)$ generators --- exactly those of $SO(2N)$. Now the 
subgroup decomposition $SO(2N) \supset SU(N) \otimes U(1)$ takes the form
\begin{equation}
\{N(2N-1)\} = \{N^2 - 1\} \oplus \{1\} \oplus \{\frac{N(N-1)}{2}\} \oplus
\{\overline{\frac{N(N-1)}{2}}\}.
\end{equation}

To summarize, in D-theory classical real and complex vectors $s$ and $z$ are 
replaced by vectors of operators $S$ and $Z$ which are embedded in $SO(N+1)$ 
and $SU(N+1)$ algebras, respectively. Similarly, classical real and complex
valued matrices $o$ and $u$ are replaced by matrices $O$ and $U$ with operator
valued elements which are embedded in the algebras of $SO(2N)$ and $SU(2N)$.
In addition, $2N \times 2N$ symplectic matrices, as well as $N \times N$ 
symmetric and anti-symmetric complex tensors are represented by the embedding 
algebras $Sp(2N)$, $Sp(N)$, and $SO(2N)$, respectively. In the next section, we
will use these basic building blocks to construct a variety of D-theory models,
both with global and with local symmetries.

\section{D-theory Formulation of Various Models}

In this section various quantum field theories are formulated in the framework 
of D-theory. Several models with various global and local symmetries are
constructed explicitly. The action operator $H$ of a model is defined on a
$d$-dimensional lattice. It has the form of a Hamiltonian that propagates the
system in an extra dimension and replaces the Euclidean action in the standard
formulation of lattice field theory. In D-theory, the partition function takes
the quantum form $Z = \mbox{Tr} \exp(- \beta H)$. The dynamics of the various 
models --- in particular, the details of their dimensional reduction --- will 
be discussed in section 5.

\subsection{$O(N)$ Quantum Spin Models}

In the standard formulation of lattice field theory, $O(N)$ models are defined
in terms of classical real $N$-component unit-vector fields $\vec s_x$. Their
action is given by
\begin{equation}
S[s] = - \sum_{x,\mu} \vec s_x \cdot \vec s_{x+\hat\mu},
\end{equation}
which is obviously invariant under $SO(N)$ rotations. In D-theory, $\vec s_x$ 
is replaced by an $N$-component vector of Hermitean operators $\vec S_x$ 
which represent $N$ of the generators of $SO(N+1)$. The corresponding action 
operator is
\begin{equation}
H = J \sum_{x,\mu} \vec S_x \cdot \vec S_{x+\hat\mu}.
\end{equation}
Now the global $SO(N)$ symmetry of the model follows from
\begin{equation}
[H,\sum_x S^{ij}_x] = 0.
\end{equation}
The rishon representation of the local $SO(N+1)$ algebra on each lattice site
$x$ is given by
\begin{equation}
S^i_x = -i (c^0_x c^i_x - c^i_x c^0_x), \
S^{ij}_x = -i (c^i_x c^j_x - c^i_x c^j_x).
\end{equation}
These ``Majorana'' rishon operators obey the anti-commutation relations
\begin{equation}
\{c^0_x,c^0_y\} = \delta_{xy}, \ \{c^0_x,c^i_y\} = 0, \ 
\{c^i_x,c^j_y\} = \delta_{xy} \delta_{ij}.
\end{equation}
Like the Heisenberg model, the $O(N)$ quantum spin model can be formulated 
using different representations --- in this case the representations of 
$SO(N+1)$. The rishon representation is just one particular choice, namely the
fundamental spinorial representation of $SO(N+1)$.

\subsection{$SO(N)$ Chiral Quantum Spin Models}

Let us consider $SO(N)_L \otimes SO(N)_R$ chiral spin models. In the standard
Wilson approach, they are formulated in terms of classical orthogonal $SO(N)$ 
matrices $o$ with an action
\begin{equation}
S[u] = - \sum_{x,\mu} \mbox{Tr} (o^\dagger_x o_{x+\hat\mu} + 
o^\dagger_{x+\hat\mu} o_x),
\end{equation}
where the dagger reduces to the transpose. The above action is invariant under 
global $SO(N)_L \otimes SO(N)_R$ transformations
\begin{equation}
o'_x = \exp(i \vec \alpha_+ \cdot \vec \lambda) o_x 
\exp(- i \vec \alpha_- \cdot \vec \lambda),
\end{equation}
where $\vec \lambda$ are the anti-Hermitean generators of $SO(N)$. 

The $SO(N)$ chiral quantum spin model of D-theory is defined by the action 
operator
\begin{equation}
H = J \sum_{x,\mu} \mbox{Tr} (O^\dagger_x O_{x+\hat\mu} +
O^\dagger_{x+\hat\mu} O_x),
\end{equation}
where $O_x$ are matrices consisting of $N^2$ Hermitean generators of $SO(2N)$.
The unitary operator generating the global $SO(N)_L \otimes SO(N)_R$ symmetry 
transformation takes the form $\exp(i \vec \alpha_+ \cdot \sum_x \vec L_x)
\exp(i \vec \alpha_- \cdot \sum_x \vec R_x)$ and
\begin{equation}
[H,\sum_x \vec L_x] = [H,\sum_x \vec R_x] = 0. 
\end{equation}
The ``Majorana'' rishon representation of the site-based $SO(2N)$ algebra is 
given by
\begin{eqnarray}
&&O^{ij}_x = -i (c_{x,+}^i c_{x,-}^j - c_{x,-}^j c_{x,+}^i), \nonumber \\
&&L^{ij}_x = -i (c_{x,+}^i c_{x,+}^j - c_{x,+}^j c_{x,+}^i), \nonumber \\ 
&&R^{ij}_x = -i (c_{x,-}^i c_{x,-}^j - c_{x,-}^j c_{x,-}^i),
\end{eqnarray}
and the anti-commutation relations take the form
\begin{equation}
\{c^i_{x,+},c^j_{y,+}\} = \{c^i_{x,-},c^j_{y,-}\} = \delta_{xy} \delta_{ij}, \
\{c^i_{x,+},c^j_{y,-}\} = 0.
\end{equation}
Using the embedding algebra $Sp(2N)$, it is straightforward to construct an
$Sp(N)$ chiral quantum spin model along the same lines.

\subsection{$U(N)$ and $SU(N)$ Chiral Quantum Spin Models}

Chiral spin models with an $SU(N)_L \otimes SU(N)_R \otimes U(1)$ global 
symmetry are usually formulated in terms of classical complex $U(N)$ matrices 
$u$ with an action
\begin{equation}
S[u] = - \sum_{x,\mu} \mbox{Tr} (u^\dagger_x u_{x+\hat\mu} + 
u^\dagger_{x+\hat\mu} u_x).
\end{equation}
This action is invariant under global $SU(N)_L \otimes SU(N)_R$ transformations
\begin{equation}
u'_x = \exp(i \vec \alpha_+ \cdot \vec \lambda) u_x 
\exp(- i \vec \alpha_- \cdot \vec \lambda),
\end{equation}
as well as under global $U(1)$ transformations
\begin{equation}
u'_x = \exp(i \alpha) u_x,
\end{equation}
which are insensitive to left or right. By restricting the matrices $u_x$ to
$SU(N)$ the symmetry of the model can be reduced to $SU(N)_L \otimes SU(N)_R$.
In D-theory the $U(N)$ chiral model is described by the action operator
\begin{equation}
H = J \sum_{x,\mu} \mbox{Tr} (U^\dagger_x U_{x+\hat\mu} +
U^\dagger_{x+\hat\mu} U_x),
\end{equation}
where $U_x$ are matrices consisting of $2N^2$ Hermitean generators of $SU(2N)$.
The unitary operator that generates a global symmetry transformation in the
Hilbert space is given by $\exp(i \vec \alpha_+ \cdot \sum_x \vec L_x)
\exp(i \vec \alpha_- \cdot \sum_x \vec R_x) \exp(i \alpha \sum_x T_x)$, 
and the $SU(N)_L \otimes SU(N)_R \otimes U(1)$ invariance of the model now 
follows from
\begin{equation}
[H,\sum_x \vec L_x] = [H,\sum_x \vec R_x] = [H,\sum_x T_x] = 0. 
\end{equation}
The rishon representation of the $SU(2N)$ algebra per local site is given by
\begin{eqnarray}
&&U^{ij}_x = c_{x,+}^{i \dagger} c_{x,-}^j, \ 
\vec L_x = \sum_{ij} c_{x,+}^{i \dagger} \vec \lambda_{ij} c_{x,+}^j, \
\vec R_x = \sum_{ij} c_{x,-}^{i \dagger} \vec \lambda_{ij} c_{x,-}^j, \nonumber
\\
&&T_x = \sum_i (c_{x,+}^{i \dagger} c_{x,+}^i - c_{x,-}^{i \dagger} c_{x,-}^i).
\end{eqnarray}
The rishon operators obey canonical anti-commutation relations
\begin{eqnarray}
&&\{c^i_{x,+},c^{j \dagger}_{y,+}\} = 
\{c^i_{x,-},c^{j \dagger}_{y,-}\} = \delta_{xy} \delta_{ij}, \ 
\{c^i_{x,+},c^{j \dagger}_{y,-}\} = 
\{c^i_{x,-},c^{j \dagger}_{y,+}\} = 0, \nonumber \\
&&\{c^i_{x,\pm},c^j_{y,\pm}\} = 0, \
\{c^{i \dagger}_{x,\pm},c^{j \dagger}_{y,\pm}\} = 0. 
\end{eqnarray}
The action operator commutes with the local rishon number
\begin{equation}
{\cal N}_x = \sum_i (c_{x,+}^{i \dagger} c_{x,+}^i + 
c_{x,-}^{i \dagger} c_{x,-}^i).
\end{equation} 
Selecting a fixed rishon number corresponds to choosing a representation of 
$SU(2N)$.

In order to reduce the symmetry of the quantum spin model from 
$SU(N)_L \otimes SU(N)_R \otimes U(1)$ to $SU(N)_L \otimes SU(N)_R$ one breaks 
the additional $U(1)$ symmetry by adding the real part of the determinant of 
each matrix to the action operator
\begin{equation}
\label{Hamdet}
H = J \sum_{x,\mu} \mbox{Tr} (U^\dagger_x U_{x+\hat\mu} + 
U^\dagger_{x+\hat\mu} U_x)
+ J' \sum_x \ (\mbox{det} U_x + \mbox{det} U^\dagger_x).
\end{equation}
One should note that the definition of $\mbox{det} U_x$ does not suffer from 
operator ordering ambiguities. In rishon representation it takes the form
\begin{eqnarray}
\mbox{det} U_x&=&\frac{1}{N!} \epsilon_{i_1 i_2 ... i_N} 
(U_x)_{i_1 i'_1} (U_x)_{i_2 i'_2} ... (U_x)_{i_N i'_N} 
\epsilon_{i'_1 i'_2 ... i'_N} \nonumber \\ 
&=&\frac{1}{N!} \epsilon_{i_1 i_2 ... i_N} 
c^{i_1}_{x,+} c^{i'_1 \dagger}_{x,-} 
c^{i_2}_{x,+} c^{i'_2 \dagger}_{x,-} ... 
c^{i_N}_{x,+} c^{i'_N \dagger}_{x,-} 
\epsilon_{i'_1 i'_2 ... i'_N} \nonumber \\
&=&N! \ c^1_{x,+} c^{1 \dagger}_{x,-} 
c^2_{x,+} c^{2 \dagger}_{x,-} ... 
c^N_{x,+} c^{N \dagger}_{x,-}.
\end{eqnarray}
Only when this operator acts on a state with ${\cal N}_x = N$ rishons it can 
give a non-vanishing result. Thus, the $U(1)$ symmetry can be eliminated via 
the determinant, only when one works with ${\cal N}_x = N$ fermionic rishons on
each site. This corresponds to choosing the $(2N)!/(N!)^2$-dimensional 
representation of $SU(2N)$ with a totally antisymmetric Young tableau with $N$ 
boxes.

\subsection{$U(N)$ and $SU(N)$ Quantum Link Models}

Wilson's formulation of lattice gauge theory uses classical complex $SU(N)$ 
parallel transporter link matrices $u_{x,\mu}$ with an action
\begin{equation}
S[u] = - \sum_{x,\mu \neq \nu} \mbox{Tr} (u_{x,\mu} u_{x+\hat\mu,\nu} 
u^\dagger_{x+\hat\nu,\mu} u^\dagger_{x,\nu}).
\end{equation}
The action is invariant under $SU(N)$ gauge transformations
\begin{equation}
u'_{x,\mu} = \exp(i \vec \alpha_x \cdot \vec \lambda) u_{x,\mu}
\exp(- i \vec \alpha_{x+\hat\mu} \cdot \vec \lambda).
\end{equation}
In D-theory, the action is replaced by the action operator
\begin{equation}
\label{Hamiltonian}
H = J \sum_{x,\mu \neq \nu} \mbox{Tr} (U_{x,\mu} U_{x+\hat\mu,\nu} 
U^\dagger_{x+\hat\nu,\mu} U^\dagger_{x,\nu}).
\end{equation}
Here the elements of the $N \times N$ quantum link operators $U_{x,\mu}$ 
consist of generators of $SU(2N)$. Gauge invariance now means that $H$ commutes
with the local generators $\vec G_x$ of gauge transformations at the site $x$, 
which obey
\begin{equation}
[G^a_x,G^b_y] = 2 i \delta_{xy} f_{abc} G^c_x.
\end{equation}
Gauge covariance of a quantum link variable requires
\begin{equation}
U'_{x,\mu} = \prod_y \exp(- i \vec \alpha_y \cdot \vec G_y) U_{x,\mu} 
\prod_z \exp(i \vec \alpha_z \cdot \vec G_z) = 
\exp(i \vec \alpha_x \cdot \vec \lambda) U_{x,\mu}
\exp(- i \vec \alpha_{x+\hat\mu} \cdot \vec \lambda),
\end{equation}
where $\prod_x \exp(i \vec \alpha_x \cdot \vec G_x)$ is the unitary operator 
that represents a general gauge transformation in Hilbert space. The above 
equation implies the following commutation relation
\begin{equation}
\label{GUcommutator}
[\vec G_x,U_{y,\mu}] = \delta_{x,y+\hat\mu} U_{y,\mu} \vec \lambda -
\delta_{x,y} \vec \lambda U_{y,\mu}.
\end{equation}
It is straightforward to show that this is satisfied when we write
\begin{equation}
\vec G_x = \sum_\mu (\vec R_{x-\hat\mu,\mu} + \vec L_{x,\mu}),
\end{equation}
where $\vec R_{x,\mu}$ and $\vec L_{x,\mu}$ are generators of right and left 
gauge transformations of the link variable $U_{x,\mu}$. The commutation 
relations of eq.(\ref{GUcommutator}) imply
\begin{equation}
[\vec R_{x,\mu},U_{y,\nu}] = 
\delta_{x,y} \delta_{\mu\nu} U_{x,\mu} \vec\lambda, \ 
[\vec L_{x,\mu},U_{y,\nu}] = - \delta_{x,y} \delta_{\mu\nu} \vec\lambda 
U_{x,\mu}.
\end{equation}
The rishon representation is given by
\begin{equation}
U^{ij}_{x,\mu} = c_{x,+\mu}^{i \dagger} c_{x+\hat\mu,-\mu}^j, \ 
\vec L_{x,\mu} = \sum_{ij} c_{x,+\mu}^{i \dagger} \vec \lambda_{ij} 
c_{x,+\mu}^j, \
\vec R_{x,\mu} = \sum_{ij} c_{x+\hat\mu,-\mu}^{i \dagger} \vec \lambda_{ij} 
c_{x+\hat\mu,-\mu}^j.
\end{equation}
The rishon operators obey canonical anti-commutation relations
\begin{equation}
\{c^i_{x,\pm\mu},c^{j \dagger}_{y,\pm\nu}\} = \delta_{xy} 
\delta_{\pm\mu,\pm\nu} \delta_{ij}, \ 
\{c^i_{x,\pm\mu},c^j_{y,\pm\nu}\} = 0, \
\{c^{i \dagger}_{x,\pm\mu},c^{j \dagger}_{y,\pm\nu}\} = 0. 
\end{equation}
The whole algebra commutes with the rishon number operator
\begin{equation}
{\cal N}_{x,\mu} = \sum_i (c^{i \dagger}_{x,+\mu} c^i_{x,+\mu} +
c^{i \dagger}_{x+\hat\mu,-\mu} c^i_{x+\hat\mu,-\mu})
\end{equation}
on each individual link. Together with the generator
\begin{equation} 
T_{x,\mu} = \sum_i (c_{x,+\mu}^{i \dagger} c_{x,+\mu}^i - 
c_{x+\hat\mu,-\mu}^{i \dagger} c_{x+\hat\mu,-\mu}^i).
\end{equation}
the above operators form the link based algebra of $SU(2N)$. One finds
\begin{equation}
[T_{x,\mu},U_{y,\nu}] = 2 \delta_{x,y} \delta_{\mu\nu} U_{x,\mu},
\end{equation}
which implies that
\begin{equation}
G_x = \frac{1}{2} \sum_\mu (T_{x-\hat\mu,\mu} - T_{x,\mu})
\end{equation}
generates an additional $U(1)$ gauge transformation, i.e.
\begin{equation}
U'_{x,\mu} = \prod_y \exp(- i \alpha_y G_y) U_{x,\mu} \prod_z 
\exp(i \alpha_z G_z) = \exp(i \alpha_x) U_{x,\mu} \exp(- i \alpha_{x+\mu}).
\end{equation}
Indeed the action operator of eq.(\ref{Hamiltonian}) is also invariant under 
the extra $U(1)$ gauge transformations and thus describes a $U(N)$ lattice 
gauge theory.

In the $SU(2)$ quantum link model constructed in \cite{Cha97} the link
matrices are represented as $O(4)$-vectors. This leads to the algebra of 
$SO(5) = Sp(2)$ instead of $SU(4)$. That construction does not contain the 
additional $U(1)$ gauge symmetry but it can not be generalized to $SU(N)$. 
Instead, its generalization to larger $N$ naturally leads to $Sp(N)$ gauge
theories, which in D-theory are embedded in the algebraic structure of 
$Sp(2N)$. In order to reduce the symmetry of the quantum link model from $U(N)$
to $SU(N)$ one breaks the additional $U(1)$ gauge symmetry by adding the real 
part of the determinant of each link matrix to the action operator
\begin{equation}
\label{Hamdetgauge}
H = J \sum_{x,\mu \neq \nu} \mbox{Tr} (U_{x,\mu} U_{x+\hat\mu,\nu} 
U^\dagger_{x+\hat\nu,\mu} U^\dagger_{x,\nu})
+ J' \sum_{x,\mu} \ (\mbox{det} U_{x,\mu} + \mbox{det} U^\dagger_{x,\mu}).
\end{equation}
The $U(N)$ symmetry can be reduced to $SU(N)$ via the determinant only when 
one works with ${\cal N}_{x,\mu} = N$ rishons on each link. Again, this 
corresponds to choosing the $(2N)!/(N!)^2$-dimensional representation of 
$SU(2N)$.

\subsection{$SO(N)$ Quantum Link Models}

Lattice gauge theory can also be formulated with real valued classical 
orthogonal parallel transporter link matrices $o_{x,\mu}$, i.e.\ with the gauge
group $SO(N)$. Again, the action takes the form
\begin{equation}
S[o] = - \sum_{x,\mu \neq \nu} \mbox{Tr} (o_{x,\mu} o_{x+\hat\mu,\nu} 
o^\dagger_{x+\hat\nu,\mu} o^\dagger_{x,\nu}),
\end{equation}
but the dagger now reduces to a transpose. In D-theory, the action is replaced 
by the action operator
\begin{equation}
H = J \sum_{x,\mu \neq \nu} \mbox{Tr} (O_{x,\mu} O_{x+\hat\mu,\nu} 
O^\dagger_{x+\hat\nu,\mu} O^\dagger_{x,\nu}).
\end{equation}
The elements of the $N \times N$ quantum link operators $O_{x,\mu}$ consist of 
generators of $SO(2N)$. Again, gauge invariance implies that $H$ commutes
with the local generators $G_x$ of gauge transformations which are given by
\begin{equation}
G^{ij}_x = \sum_\mu (R^{ij}_{x-\hat\mu,\mu} + L^{ij}_{x,\mu}).
\end{equation}
The rishon representation takes the form
\begin{eqnarray}
&&O^{ij}_{x,\mu} = -i (c_{x,+\mu}^i c_{x+\hat\mu,-\mu}^j -
c_{x+\hat\mu,-\mu}^j c_{x,+\mu}^i), \nonumber \\
&&L^{ij}_{x,\mu} = -i (c_{x,+\mu}^i c_{x,+\mu}^j -
c_{x,+\mu}^j c_{x,+\mu}^i), \nonumber \\ 
&&R^{ij}_{x,\mu} = -i (c_{x+\hat\mu,-\mu}^i c_{x+\hat\mu,-\mu}^j -
c_{x+\hat\mu,-\mu}^j c_{x+\hat\mu,-\mu}^i).
\end{eqnarray}
The ``Majorana'' rishon operators obey the anti-commutation relations
\begin{equation}
\{c^i_{x,\pm\mu},c^j_{y,\pm\nu}\} = 
\delta_{xy} \delta_{\pm\mu,\pm\nu} \delta_{ij}.
\end{equation}

\subsection{Quantum Link Models with Other Gauge Groups}

It is interesting to ask how gauge theories with other gauge groups can be
formulated in D-theory. Besides the Lie groups $SU(N)$ and $SO(N)$ there
is also the symplectic group $Sp(N)$. Quantum link models with the symplectic 
gauge group $Sp(N)$ can be constructed in complete analogy to the $SU(N)$ and 
$SO(N)$ cases. As discussed in section 3.5, the embedding algebra is then 
$Sp(2N) \supset Sp(N)_L \otimes Sp(N)_R$. In particular, the construction of 
the $Sp(1) = SU(2)$ quantum link model uses the embedding algebra 
$Sp(2) = SO(5)$. This case was discussed in detail in \cite{Cha97}.

Besides the main sequence Lie groups $SU(N)$, $SO(N)$, and $Sp(N)$ there are
also the exceptional Lie groups $G(2)$, $F(4)$, $E(6)$, $E(7)$, and $E(8)$.
These groups must be treated on a case by case basis. Here we only consider the
simplest exceptional group $G(2)$ which is a subgroup of $SO(7)$. The group
$G(2)$ has rank 2, it has 14 generators, and it contains those $SO(7)$ matrices
$o$ that satisfy the cubic constraint
\begin{equation}
\label{cubic}
T_{ijk} = T_{lmn} o^{li} o^{mj} o^{nk}.
\end{equation}
Here $T$ is a totally anti-symmetric tensor whose non-zero elements follow by 
anti-symmetrization from
\begin{equation}
\label{tensor}
T_{127} = T_{154} = T_{163} = T_{235} = T_{264} = T_{374} = T_{576} = 1.
\end{equation}
Eq.(\ref{tensor}) implies that eq.(\ref{cubic}) represents 7 non-trivial
constraints which reduce the 21 degrees of freedom of $SO(7)$ to the 14 
parameters of $G(2)$. In D-theory one needs $14 + 14 = 28$ generators of the
embedding algebra in order to represent the $G(2)_L \otimes G(2)_R$ gauge 
symmetry at the two ends of a link. In addition, one needs 49 Hermitean 
generators to represent the $7 \times 7$ real matrix elements of a link 
variable. Hence, altogether one needs at least $28 + 49 = 77$ generators. It is
tempting to try to embed this structure in the algebra $E(6)$ which has 78 
generators. However, this does not work because $E(6)$ does not even contain 
$G(2)_L \otimes G(2)_R$ as a subgroup \cite{Sla81}. Instead, one can use the 
embedding algebra $SO(14)$ which is used in the $SO(7)$ quantum link model. The
group $SO(14)$ has 91 generators. Under the subgroup decomposition 
$SO(14) \supset G(2)_L \otimes G(2)_R$ they decompose as
\begin{equation}
\{91\} = \{14,1\} \oplus \{7,1\} \oplus \{1,14\} \oplus \{1,7\} \oplus \{7,7\}.
\end{equation}
If one simply uses the action operator of the $SO(7)$ quantum link model, the
theory is by construction $SO(7)$ --- and not just $G(2)$ --- gauge invariant.
In order to explicitly break the symmetry down to $G(2)$ one adds another term 
to the action operator
\begin{equation}
H = J \sum_{x,\mu \neq \nu} \mbox{Tr} (O_{x,\mu} O_{x+\hat\mu,\nu} 
O^\dagger_{x+\hat\nu,\mu} O^\dagger_{x,\nu}) + 
J' \sum_{x,\mu} T_{ijk} T_{lmn} O^{li}_{x,\mu} O^{mj}_{x,\mu} O^{nk}_{x,\mu}.
\end{equation}
The additional term is $G(2)$ but not $SO(7)$ gauge invariant. This procedure
is similar to the way in which the gauge symmetry of a $U(N)$ quantum link 
model was reduced to $SU(N)$.

\subsection{Quantum $CP(N-1)$ Models}

$CP(N-1)$ models are interesting because they have a global $SU(N)$ symmetry 
as well as a $U(1)$ gauge invariance. However, in this case, the gauge fields 
are just auxiliary fields. In the standard formulation of lattice field 
theory, $CP(N-1)$ models are formulated in terms of classical complex 
unit-vectors $z_x$ and complex link variables $u_{x,\mu} \in U(1)$. The action 
can be written as
\begin{equation}
S[z,u] = - \sum_{x,\mu} (z^\dagger_x u_{x,\mu} z_{x+\hat\mu} +
z^\dagger_{x+\hat\mu} u^\dagger_{x,\mu} z_x).
\end{equation}
Note that there is no plaquette term for the gauge field. Consequently, the
gauge field can be integrated out --- it only acts as an auxiliary field. In
D-theory, the corresponding action operator takes the form
\begin{equation}
H = J \sum_{x,\mu} (Z^\dagger_x U_{x,\mu} Z_{x+\hat\mu} +
Z^\dagger_{x+\hat\mu} U^\dagger_{x,\mu} Z_x).
\end{equation}
Here the $N$ components $Z^i_x$ of the quantum spin $Z_x$ consist of $2N$ 
Hermitean generators of $SU(N+1)$ and $U_{x,\mu} = S^1_{x,\mu} + 
i S^2_{x,\mu} = S^+_{x,\mu}$ is the raising operator of an $SU(2)$ algebra on 
each link. In rishon representation we have
\begin{equation}
Z^i_x = c^{0 \dagger}_x c^i_x, \ 
U_{x,\mu} = c^\dagger_{x,+\mu} c_{x+\hat\mu,-\mu}.
\end{equation}
Global $SU(N)$ transformations are generated by
\begin{equation}
\vec G = \sum_x \sum_{ij} c^{i \dagger}_x \vec \lambda_{ij} c^j_x,
\end{equation}
and the generator of $U(1)$ gauge transformations takes the form
\begin{equation}
G_x = \frac{1}{2} \sum_\mu (T_{x-\hat\mu,\mu} - T_{x,\mu}) + 
\sum_i c^{i \dagger}_x c^i_x.
\end{equation}
In rishon representation we have
\begin{equation}
T_{x,\mu} = c^\dagger_{x,+\mu} c_{x,+\mu} - 
c^\dagger_{x+\hat\mu,-\mu} c_{x+\hat\mu,-\mu}.
\end{equation}
The invariance properties of the model follow from $[H,\vec G] = [H,G_x] = 0$.
In this case, the rishon numbers on each site and on each link are separately
conserved.

It should be clear by now that D-theory is a very general structure which
naturally provides us with lattice field theories formulated in terms of 
discrete variables --- quantum spins and quantum links. The cases worked out
here in some detail are examples that demonstrate the generality of D-theory. 
There are certainly more models one could investigate. In particular, it is
straightforward to include fermion fields \cite{Bro99}. In fact, the additional
dimension of D-theory provides a natural framework for domain wall fermions.
Next, we will argue that D-theory does not define a new set of field theories. 
It just provides a new non-perturbative regularization and quantization of the 
corresponding classical models. In order to understand this, we now address the
issue of dimensional reduction.

\section{Classical Fields from Dimensional Reduction of Discrete Variables}

As we have seen in detail in section 2, the quantum Heisenberg model provides
a D-theory regularization for the $d$-dimensional $O(3)$ model. The connection 
between the two models is established using chiral perturbation theory. The 
Goldstone boson fields $\vec s$ describing the low-energy dynamics of the 
Heisenberg 
model emerge as collective excitations of the quantum spins. By dimensional 
reduction they become the effective 2-d fields of a lattice $O(3)$ model. The 
continuum limit of this lattice model is reached when the extent $\beta$ of the
extra Euclidean dimension becomes large. The success of D-theory relies 
entirely on the fact that the $(d+1)$-dimensional theory is mass\-less, i.e.\ 
that in the ground state of the quantum Heisenberg model the $SO(3)$ symmetry 
is spontaneously broken to $SO(2)$. Only then chiral perturbation theory 
applies and dimensional reduction occurs. The same is true for other D-theory 
quantum spin models. In case of the spin 1/2 quantum Heisenberg model it 
required very precise numerical simulations before one could be sure that 
spontaneous symmetry breaking indeed occurs. For the other D-theory spin models
similar simulations have not yet been done. In the following discussion of the 
dynamics we will assume that in D-theory the same symmetry breaking pattern 
arises as in the corresponding Wilsonian lattice field theory. This can indeed
be shown for sufficiently large representations of the corresponding embedding
algebra \cite{Sch01}. In practical numerical calculations one would like to 
work with small representations. We expect that efficient cluster algorithms 
can be constructed for various D-theory models. Hence, it should be possible to
investigate if massless modes indeed exist in the $(d+1)$-dimensional theory,
and thus perform detailed tests of the dynamical picture that is drawn here.

\subsection{$O(N)$ Models}

Let us investigate the $O(N)$ quantum spin models constructed in the previous 
section. First, we consider $d = 2$. The quantum statistical partition function
$Z = \mbox{Tr} \exp(- \beta H)$ can then be represented as a 
$(2+1)$-dimensional path integral with finite extent $\beta$ in the extra
Euclidean dimension. At $\beta = \infty$ we have an infinite 3-d system with an
$SO(N)$ symmetry. In Wilson's formulation of lattice field theory, such models
can be in a mass\-less phase in which the symmetry is spontaneously broken to
$SO(N-1)$. Here we assume that the same is true for the quantum spin model. 
Just as in the Heisenberg model, for small representations of the embedding 
algebra $SO(N+1)$, this assumption can only be tested in numerical simulations.
First, we limit ourselves to the non-Abelian $N > 2$ case. The Abelian $N = 2$ 
case of the quantum XY model requires a separate discussion. When an $SO(N)$ 
symmetry breaks spontaneously down to $SO(N-1)$, the low-energy Goldstone boson
dynamics is described by chiral perturbation theory. In this case, the 
Goldstone bosons are represented by fields in the coset space
$SO(N)/SO(N-1) = S^{N-1}$ which is an $(N-1)$-dimensional sphere. Consequently,
the Goldstone boson fields are $N$-component unit vectors $\vec s$ and their 
low-energy effective action is given by
\begin{equation}
S[\vec s] = \int_0^\beta dx_3 \int d^2x \ \frac{\rho_s}{2}
\left(\partial_\mu \vec s \cdot \partial_\mu \vec s + 
\frac{1}{c^2} \partial_3 \vec s \cdot \partial_3 \vec s\right).
\end{equation}
Again, $c$ and $\rho_s$ are the spin wave velocity and the spin stiffness.

Dimensional reduction now occurs exactly as in the $O(3)$ Heisenberg model.
When the extent $\beta$ of the extra dimension becomes finite, the correlation
length $\xi$ is much larger than $\beta$ and the system becomes 2-dimensional.
In two dimensions, however, the Mermin-Wagner-Coleman theorem implies that the
Goldstone bosons pick up a mass. Hasenfratz and Niedermayer have determined the
corresponding finite correlation length \cite{Has91}
\begin{equation}
\xi = \beta c \left(\frac{e(N-2)}{16 \pi \beta \rho_s}\right)^{1/(N-2)} 
\Gamma\left(1 + \frac{1}{N - 2}\right) 
\exp\left(\frac{2 \pi \beta \rho_s}{N - 2}\right)
\left[1 - \frac{1}{4 \pi \beta \rho_s} + {\cal O}(1/\beta^2 \rho_s^2)\right].
\end{equation}
Again, $1/g^2 = \beta \rho_s$ defines the coupling constant $g$ of an effective
$O(N)$ symmetric Wilsonian lattice gauge theory with lattice spacing $\beta c$.
The exponential divergence of $\xi$ is due to asymptotic freedom. In this case,
the 1-loop $\beta$-function coefficient is $(N-2)/2 \pi$. The continuum limit 
of the effective lattice model is reached in the $g \rightarrow 0$ limit, i.e.\
when the extent $\beta$ of the extra dimension becomes large. Still, in 
physical units of $\xi$, this extent becomes negligible and the 
$(2+1)$-dimensional D-theory model is reduced to the 2-d $O(N)$ model.

In higher dimensions the situation is exactly as in the $d$-dimensional 
Heisenberg model. Assuming again that the ground state has a broken symmetry,
dimensional reduction occurs once $\beta$ becomes finite. However, in contrast
to the $d = 2$ case, the Goldstone bosons remain mass\-less after dimensional
reduction. Still, when $\beta$ becomes too small, one enters the symmetric 
phase with a finite correlation length. The universal continuum behavior at the
corresponding second order phase transition is naturally contained in the
framework of D-theory. For general $N$, the $d = 1$ case is different from 
$N = 3$ because there are no instantons --- and thus no $\theta$-term --- for 
$N \neq 3$. Consequently, we expect the 1-d $O(N)$ quantum spin chain to be
always massive. In that case, dimensional reduction would not occur and no 
universal continuum physics arises.

Let us now turn to the $N = 2$ case of the 2-d quantum XY model. The classical 
2-d XY model has a Kosterlitz-Thouless transition separating a massive phase
with a vortex condensate from a mass\-less spin wave phase. In this case, the
Mermin-Wagner-Coleman theorem is evaded because the spin waves do not interact
with each other. There is numerical evidence for a Kosterlitz-Thouless 
transition at finite $\beta$ also in the quantum XY model \cite{Din90,Har97}. 
For a large extent of the third Euclidean dimension, one is in a mass\-less 
phase with an infinite correlation length, which means that dimensional 
reduction occurs already at finite $\beta$. Hence, the dimensional reduction in
the 2-d quantum XY model is analogous to the behavior of $O(N)$ models with 
$N \geq 3$ in dimensions $d \geq 3$. The continuum field theory resulting from 
the 2-d quantum XY model is the 2-d classical XY model --- an $SO(2)$-symmetric
theory of a free mass\-less scalar field. 

\subsection{$SU(N)$, $Sp(N)$, and $SO(N)$ Chiral Models}

Let us consider the $d$-dimensional $SU(N)$ chiral quantum spin model. Its
partition function is described by a $(d+1)$-dimensional path integral with an 
extent $\beta$ of the extra dimension. At $\beta = \infty$ we have a 
$(d+1)$-dimensional system with an $SU(N)_L \otimes SU(N)_R$ chiral symmetry.
For a sufficiently large representation of the embedding algebra $SU(2N)$ it
has been shown that the quantum chiral spin model has the same spontaneous 
symmetry breaking pattern that occurs in Wilson's lattice field theory
\cite{Sch01}. Consequently, the $SU(N)_L \otimes SU(N)_R$ symmetry breaks down
to $SU(N)_{L=R}$ and the coset space for the Goldstone boson fields is 
$SU(N)_L \otimes SU(N)_R/SU(N)_{L=R} = SU(N)$. For $d = 2$ the low-energy 
effective action takes the form
\begin{equation}
S[U] = \int_0^\beta dx_3 \int d^2x \ \frac{\rho_s}{4} \mbox{Tr}
\left(\partial_\mu U^\dagger \partial_\mu U + 
\frac{1}{c^2} \partial_3 U^\dagger  \partial_3 U\right).
\end{equation}
The dimensional reduction is exactly like in $O(N)$ models with $N \geq 3$. In 
particular, in $d = 2$, due to the Mermin-Wagner-Coleman theorem, one again
obtains a finite correlation length $\xi$. Using the exact mass gap of the 2-d 
$SU(N)$ chiral model \cite{Bal92}, up to order $1/\beta \rho_s$ corrections, 
one obtains
\begin{equation}
\xi = \beta c \left(\frac{e}{N \beta \rho_s}\right)^{1/2} 
\frac{\exp(2 \pi \beta \rho_s/N)}{4 \sin(\pi/N)} 
\left[1 + {\cal O}(1/\beta \rho_s)\right].
\end{equation}
Note that this is consistent with the $O(4)$ model result for $N = 2$. Again, 
due to asymptotic freedom, dimensional reduction occurs in the 
$\beta \rightarrow \infty$ continuum limit. In higher dimensions dimensional 
reduction already occurs at finite $\beta$. 

Similarly, using the exact mass gap of the 2-d $Sp(N)$ and $SO(N)$ chiral
models \cite{Hol94} one obtains corresponding results for the correlation
length. For the $Sp(N)$ chiral spin models one gets
\begin{equation}
\xi = \beta c \left(\frac{e}{(N+1) \beta \rho_s}\right)^{1/2} 
\frac{\exp(2 \pi \beta \rho_s/(N+1))}{2^{(3N+1)/(N+1)} \sin(\pi/(N+1))} 
\left[1 + {\cal O}(1/\beta \rho_s)\right].
\end{equation}
This is consistent with the $SU(2) \otimes SU(2)$ model for $N = 1$.

For $SO(N)$ with $N \geq 7$ the correlation length is given by
\begin{equation}
\xi = \beta c \left(\frac{e}{(N-2) \beta \rho_s}\right)^{1/2}
\frac{\exp(2 \pi \beta \rho_s/(N-2))}{2^{(2N-2)/(N-4)} \sin(\pi/(N-2))} 
\left[1 + {\cal O}(1/\beta \rho_s)\right].
\end{equation}
Since $SO(3) \simeq SU(2) = Sp(1)$, $SO(5) \simeq Sp(2)$, and 
$SO(6) \simeq SU(4)$, the cases $N = 3$, 5, and 6 are covered by the
corresponding $SU(N) \otimes SU(N)$ and $Sp(N) \otimes Sp(N)$ chiral models.
Since $SO(4) \simeq SU(2) \otimes SU(2)$ the $N = 4$ case corresponds to two
decoupled $SU(2) \otimes SU(2)$ chiral models.

\subsection{$SU(N)$, $Sp(N)$, $SO(N)$, and $U(1)$ Gauge Theories}

Let us consider $SU(N)$ non-Abelian gauge theories, first in $d = 4$. The 
action operator of the corresponding quantum link model, which is defined on 
a 4-d lattice, describes the evolution of the system in a fifth Euclidean
direction. The partition function $Z = \mbox{Tr} \exp(- \beta H)$ can then be
represented as a $(4+1)$-d path integral. Note that we have not included a
projector on gauge invariant states, i.e.\ gauge variant states also propagate
in the fifth direction. This means that we do not impose a Gauss law in the
unphysical direction. Not imposing Gauss' law implies $A_5 = 0$ for the fifth 
component of the gauge potential. This is convenient, because it leaves us with
the correct field content after dimensional reduction. Of course, the physical 
Gauss law is properly imposed because the model does contain non-trivial 
Polyakov loops in the Euclidean time direction. 

Dimensional reduction in quantum link models works differently than for quantum
spins. In the spin case the spontaneous breakdown of a global symmetry provides
the mass\-less Goldstone modes that are necessary for dimensional reduction. On
the other hand, when a gauge symmetry breaks spontaneously, the Higgs mechanism
gives mass to the gauge bosons and dimensional reduction would not occur.
Fortunately, non-Abelian gauge theories in five dimensions are generically in
a mass\-less Coulomb phase \cite{Cre79}. This has recently been verified in 
detail for 5-d $SU(2)$ and $SU(3)$ lattice gauge theories using Wilson's 
formulation \cite{Bea98a}. For sufficiently large representations of the 
embedding algebra $SU(2N)$ the same is true for quantum link models 
\cite{Sch01}. Whether a $(4+1)$-d $SU(N)$ quantum link model using a small 
representation of $SU(2N)$ is still in the Coulomb phase can only be checked in
numerical simulations. The leading terms in the low-energy effective action of 
5-d Coulombic gluons take the form
\begin{equation}
S[A] = \int_0^\beta dx_5 \int d^4x \ \frac{1}{2 e^2} \left(\mbox{Tr} \ 
F_{\mu\nu} F_{\mu\nu} + \frac{1}{c^2} \mbox{Tr} \ 
\partial_5 A_\mu \partial_5 A_\mu\right).
\end{equation}
The quantum link model leads to a 5-d gauge theory characterized by the 
``velocity of light'' $c$. Note that here $\mu$ runs over 4-d indices only. The
dimensionful 5-d gauge coupling $1/e^2$ is the analog of $\rho_s$ in the spin
models. At finite $\beta$ the above theory has only a 4-d gauge invariance, 
because we have fixed $A_5 = 0$, i.e.\ we have not imposed the Gauss law. At
$\beta = \infty$ we are in the 5-d Coulomb phase with mass\-less gluons and 
thus with an infinite correlation length $\xi$. When $\beta$ is made finite, 
the extent of the extra dimension is negligible compared to $\xi$. Hence, the 
theory appears to be dimensionally reduced to four dimensions. Of course, in 
four dimensions the confinement hypothesis suggests that gluons are no longer 
mass\-less. Indeed, as it was argued in \cite{Cha97}, a finite correlation 
length
\begin{equation}
\label{SUNgap}
\xi \propto \beta c \left(\frac{11 e^2 N}{48 \pi^2 \beta}\right)^{51/121}
\exp\left(\frac{24 \pi^2 \beta}{11 N e^2}\right)
\end{equation}
is expected to be generated non-perturbatively. Here $11 N/48 \pi^2$ is the
1-loop $\beta$-function coefficient of $SU(N)$ gauge theory. In contrast to the
spin models, the exact value for the mass gap (and hence the prefactor of the
exponential) is not known in this case. For large $\beta$ the gauge coupling of
the dimensionally reduced 4-d theory is given by
\begin{equation}
1/g^2 = \beta/e^2.
\end{equation}
Thus the continuum limit $g \rightarrow 0$ of the 4-d theory is approached when
one sends the extent $\beta$ of the fifth direction to infinity. Hence,
dimensional reduction occurs when the extent of the fifth direction becomes 
large. This is due to asymptotic freedom, which implies that the correlation 
length grows exponentially with $\beta$. As in the spin models, it is useful to
think of the dimensionally reduced 4-d theory as a Wilsonian lattice theory 
with lattice spacing $\beta c$ (which has nothing to do with the lattice
spacing of the quantum link model). In fact, one can again imagine performing 
a block renormalization group transformation that averages the 5-d field over 
cubic blocks of size $\beta$ in the fifth direction and of size $\beta c$ in 
the four physical space-time directions. The block centers then form a 4-d 
space-time lattice of spacing $\beta c$ and the effective theory of the block 
averaged 5-d field is indeed a Wilsonian 4-d lattice gauge theory.

Using the 1- and 2-loop $\beta$-function coefficients, it is straightforward to
derive similar formulas for the correlation lengths of $Sp(N)$ and $SO(N)$ 
quantum link models. In the $Sp(N)$ case, one obtains
\begin{equation}
\xi \propto \beta c \left(\frac{11 e^2 (N+1)}{48 \pi^2 \beta}\right)^{51/121}
\exp\left(\frac{24 \pi^2 \beta}{11 (N+1) e^2}\right),
\end{equation}
while for $SO(N)$ with $N \geq 4$
\begin{equation}
\xi \propto \beta c \left(\frac{11 e^2 (N-2)}{48 \pi^2 \beta}\right)^{51/121}
\exp\left(\frac{24 \pi^2 \beta}{11 (N-2) e^2}\right).
\end{equation}
Since $SO(3) \simeq SU(2)$, the $N = 3$ case is covered by eq.(\ref{SUNgap}).

Let us also consider the $d = 3$ case. Then, due to confinement, there is no 
mass\-less phase in $d + 1 = 4$ dimensions. Hence, one then expects no 
dimensional reduction and no universal continuum behavior. This situation is 
analogous to the behavior of 1-d quantum spin chains discussed earlier. Still, 
in accordance with Haldane's conjecture, the $O(3)$ spin chain displays 
universal behavior because half-integer spins correspond to the mass\-less 
2-d $O(3)$ model at $\theta = \pi$. Similarly, it is possible that 3-d $SU(N)$ 
quantum link models, formulated with an appropriate representation of $SU(2N)$,
correspond to a 4-d Yang-Mills theory at $\theta \neq 0$. Again, such a theory 
may be mass\-less. Cluster algorithms for quantum link models would be an ideal
tool to investigate $\theta$-vacua in gauge theories because the corresponding 
D-theory model may not have a complex action problem.

The situation for 4-d $U(1)$ gauge theory is analogous to the 2-d XY model. In 
contrast to non-Abelian gauge theories, after dimensional reduction from five 
to four dimensions, there is no reason for the photons to pick up a mass. They 
can exist in a 4-d Coulomb phase because they are not confined. Hence, 
dimensional reduction occurs already at a finite extent of the fifth dimension 
--- not only in the $\beta \rightarrow \infty$ limit. Still, when $\beta$ 
becomes too small, one enters the strong coupling confined phase which has a 
finite correlation length. If the phase transition between the confined phase 
and the Coulomb phase is of second order one obtains universal continuum 
behavior via dimensional reduction by approaching the phase transition in the 
confined phase. For $d = 3$, a $U(1)$ quantum link model can undergo 
dimensional reduction from four to three dimensions because 4-d $U(1)$ gauge 
fields can exist in a mass\-less Coulomb phase. However, after dimensional 
reduction the correlation length becomes finite because 3-d $U(1)$ gauge 
theories are always in the confined phase \cite{Pol77,Goe81}. In fact, just as 
in non-Abelian gauge theories in four dimensions, an exponentially large 
correlation length arises. Therefore, dimensional reduction now occurs in the 
$\beta \rightarrow \infty$ limit. Hence, the universal continuum behavior of 
$U(1)$ gauge theories, both in three and four dimensions, is naturally 
contained in the framework of D-theory.

\subsection{$CP(N-1)$ Models}

First, let us consider quantum $CP(N-1)$ models in $d = 2$. Their partition 
function is again given by a $(2+1)$-d path integral with a finite extent
$\beta$ in the extra dimension. As before, in the $\beta \rightarrow \infty$ 
limit we have a 3-d system, in this case with a global $SU(N)$ and a local 
$U(1)$ symmetry. Again, assuming the same symmetry breaking pattern as in
Wilson's theory, such a system has a broken symmetry phase with only a global 
$U(N-1)$ symmetry left intact. As a consequence, the corresponding Goldstone
bosons live in the coset space $SU(N)/U(N-1) = S^{2N-1}/S^1$ --- i.e.\ they are
just the classical $N$-component complex unit-vectors $z$ of a $CP(N-1)$ model 
with a $U(1)$ gauge symmetry. Therefore, the low-energy effective action takes 
the form
\begin{equation}
S[z,A] = \int_0^\beta dx_3 \int d^2x \ \frac{\rho_s}{2}
\left[|(\partial_\mu + A_\mu) z|^2 + \frac{1}{c^2} |\partial_3 z|^2\right].
\end{equation}
Note that there is no kinetic term for the $U(1)$ gauge field $A_\mu$ --- it is
just an auxiliary field. Since we have not imposed Gauss' law for states 
propagating in the extra dimension, the third component $A_3$ of the gauge 
field vanishes. Consequently, an ordinary (not a covariant) derivative arises 
in the last term. When the above system of Goldstone bosons is considered at a 
finite extent $\beta$, it undergoes dimensional reduction to two dimensions. 
Again, due to asymptotic freedom an exponentially large correlation length
\begin{equation}
\xi \propto \beta c \left(\frac{N}{4 \pi \beta \rho_s}\right)^{2/N}
\exp\left(\frac{4 \pi \beta \rho_s}{N}\right)
\end{equation}
is generated, and dimensional reduction occurs in the $\beta \rightarrow 
\infty$ continuum limit. In this case, the exact mass gap is not known
analytically. As usual, $N/4 \pi$ is the 1-loop coefficient of the perturbative
$\beta$-function. In higher dimensions the reduction is exactly as in $O(N)$ 
models with $N \geq 3$ or in chiral models.

One-dimensional $CP(N-1)$ quantum spin chains are interesting because they
may be similar to the Heisenberg chain. Note that, at the classical level,
$CP(1) = O(3)$. In contrast to $O(N)$ models, $CP(N-1)$ models have instantons
and hence a $\theta$-angle for all $N$. This suggests to extend Haldane's
conjecture to these models. When formulated with appropriate representations of
$SU(N+1)$ (the embedding algebra of quantum $CP(N-1)$ models), quantum 
$CP(N-1)$ chains may correspond to 2-d classical $CP(N-1)$ models with a 
$\theta$-term. Still, unlike in the $O(3)$-model, this would not necessarily
mean that 1-d quantum $CP(N-1)$ chains undergo dimensional reduction. In fact,
for 2-d classical $CP(N-1)$ models one expects a first order phase transition 
at $\theta = \pi$ \cite{Sei84,Aff91}. In that case, the correlation length 
remains finite and dimensional reduction does not occur.

\section{Conclusions}

We have seen that D-theory provides a rich structure which allows us to 
formulate quantum field theories in terms of discrete variables --- quantum 
spins or quantum links. Dimensional reduction of discrete variables is a 
generic phenomenon. In $(d+1)$-dimensional quantum spin models with $d \geq 2$,
it occurs because of spontaneous symmetry breaking, while in 
$(4+1)$-dimensional non-Abelian quantum link models it is due to the presence 
of a 5-d mass\-less Coulomb phase. The inclusion of fermions is very natural 
when one follows Shamir's variant \cite{Sha93} of Kaplan's domain wall fermion 
proposal \cite{Kap92}. In particular, the fine-tuning problem of Wilson 
fermions is solved very elegantly by going to five dimensions. 

It is remarkable that D-theory treats bosons and fermions on an equal footing.
Both are formulated in a finite Hilbert space per site, both require the 
presence of an extra dimension, and both naturally have exponentially large
correlation lengths after dimensional reduction. The discrete nature of the
fundamental variables makes D-theory attractive, both from an analytic and from
a computational point of view. On the analytic side, the discrete variables
allow us to rewrite the theory in terms of fermionic rishon constituents of the
bosonic fields. This may turn out to be useful when one studies the large $N$ 
limit of various models \cite{Bae02}. In particular, one can now carry over 
powerful techniques developed for condensed matter systems (like the quantum 
Heisenberg model) to particle physics. This includes the use of very efficient 
cluster algorithms which has the potential of dramatically improving numerical 
simulations of lattice field theories. 

In D-theory the classical fields of ordinary quantum field theory arise via 
dimensional reduction of discrete variables. This requires specific dynamics 
--- namely a mass\-less theory in one more dimension. In general, the
verification of this basic dynamical ingredient of D-theory requires 
non-perturbative insight --- for example, via numerical simulations or via the 
large $N$ limit. Thus, the connection to ordinary field theory methods --- in
particular, to perturbation theory --- is rather indirect. This could be viewed
as a potential weakness, for example, because it seems hopeless to do 
perturbative QCD calculations in the framework of D-theory. However, the large
separation from perturbative methods may actually turn out to be a major 
strength of D-theory. The fact, that perturbative calculations are difficult, 
may imply that non-perturbative calculations are now easier. After all, 
D-theory provides an additional non-perturbative microscopic structure 
underlying Wilson's lattice theory. Our hope is that this structure will help
us to better understand the non-perturbative dynamics of quantum field 
theories.
 
\section*{Acknowledgements}

We are indebted to P.~Forgacs, F.~Niedermayer, and M.~Reuter for very 
interesting discussions. This work was supported in part by funds provided by 
the U.S. Department of Energy (D.O.E.) under cooperative research agreement
DE-FG02-96ER40945, by the Schweizerischer Nationalfond (SNF), and by the 
European Community's Human Potential Programme under contract
HPRN-CT-2000-00145.

\end{document}